\documentclass[]{aa}
\usepackage{graphicx}

\newcommand{\grp}    {${\rlap.}^{\circ}$}
\newcommand{\pri}    {${\rlap.}^{\prime \prime}$}
\newcommand{\rl}     {${\rlap.}^{s}$}

\newcommand{\ltsima} {$\; \buildrel < \over \sim \;$}
\newcommand{\simlt}  {\lower.5ex\hbox{\ltsima}}            
\newcommand{\gtsima} {$\; \buildrel > \over \sim \;$}
\newcommand{\simgt}  {\lower.5ex\hbox{\gtsima}}            
\newcommand{\ks}     {KS~1741$-$293}

\begin{document}

\title{Chandra X-ray counterpart of KS~1741$-$293}

\author{
J. Mart\'{\i}\inst{1,3}, J.~A. Combi\inst{1,3}, D. P\'erez-Ram\'{\i}rez\inst{1,3},J.~L. Garrido\inst{1,3}, P. Luque-Escamilla\inst{2, 3}, A.~J. Mu\~noz-Arjonilla\inst{3}, J.~R. S\'anchez-Sutil\inst{3}
}

\offprints{J. Mart\'{\i}}

\institute{
Departamento de F\'{\i}sica, EPS,
Universidad de Ja\'en, Campus Las Lagunillas s/n, A3, 23071 Ja\'en, Spain \\
\email{jmarti@ujaen.es, jcombi@ujaen.es, dperez@ujaen.es, jlg@ujaen.es}
\and Dpto. de Ing. Mec\'anica y Minera, EPS,
Universidad de Ja\'en, Campus Las Lagunillas s/n, A3, 23071 Ja\'en, Spain \\
\email{peter@ujaen.es}
\and
Grupo de Investigaci\'on FQM-322,
Universidad de Ja\'en, Campus Las Lagunillas s/n, A3, 23071 Ja\'en, Spain \\
\email{f72muara@uco.es, jrssutil@hotmail.com}
}

\date{Received / Accepted}

\authorrunning{Mart\'{\i} et al.}

\abstract
{}
{We aim to investigate the nature of the high energy source KS~1741$-$293
by revisiting the radio and infrared associations proposed in the early 1990s.
}
{Our work is mostly based on the analysis of modern survey and archive data, including
the NRAO, MSX, 2MASS and Chandra archives, and catalogues. We also have obtained deep CCD optical observations
by ourselves.}
{The coincidence of \ks\ with an extended radio and far-infrared source, tentatively suggested in 1994,
is no longer supported by modern observational data. Instead,
a Chandra source is the only peculiar object found to be consistent with all high-energy
error circles of \ks\ and we propose it to be its most likely X-ray counterpart.
We also report the existence of a non-thermal radio nebula in the vicinity of the \ks\ position with
the appearance of a supernova remnant. The possibility of being associated to this X-ray binary is discussed.
 }
{}

\maketitle

\section{Introduction}

\ks\ ($l^{II}=359$\grp 56; $b^{II}=-0$\grp 08) is a X-ray transient
first detected in 1989 by the X-ray widefield camera TTM
in the KVANT module of the Mir space station (\cite{inz1991}) and located with arc-minute accuracy.
The KVANT position had some degree of overlap with the X-ray sources MXB~1742$-$29 and MXB~1743$-$29, detected in 1976 by the
Third Small Astronomy Satellite (SAS-3).
At the time of its discovery, \ks\ was also considered as a bursting source based on two burst events with single peak structure.
The bursting nature of the source strongly pointed to a neutron star low-mass X-ray binary (LMXB),
and it appears to be classified as such in the \cite{liu01} catalogue.

\ks\ has been observed by different satellites over the years and
its detection history is summarized in Table \ref{Detections}.
While most space observatories agree about the fact that this is a
highly absorbed source ($N_H \simeq 10^{23}$ cm$^{-2}$), its
position has unfortunately remained poor at the \ltsima 1
arc-minute level. For instance, the BeppoSAX detection
(\cite{sid1999}) was reported using not its own data but the more
accurate original KVANT position. All detections are with a
comparable flux and power-law index (when available), thus
reinforcing the idea that the observed source is actually the
same.

The optical, infrared, and radio counterparts of \ks\
were searched for and investigated by \cite{ch1994}. Among possible identifications, they
tentatively quoted the extended radio G359.54$-$068 and far-infrared FIR5
sources, as well as IRAS 17417$-$2919. However, no firm candidate resulted from their work
and this remains unchanged since then.

The fact that \ks\ is today known to be among the high energy
sources dominating the Galactic center sky in hard X-ray/soft
$\gamma$-rays (\cite{bel2006}), strongly calls
for the revision of these associations proposed more than one
decade ago. This goal is at present a much more feasible task
thanks to the availability of modern radio, infrared, and X-ray
surveys with excellent sensitivity and angular resolution. For
instance, the early suggestion of possible radio/infrared
coincidences were based on single telescopes with arc-minute
angular resolution.

In this paper, we use improved multi-wavelength
data, free from the severe confusion problems of previous work,
allowing us to propose an arc-second accurate location for the
\ks\ counterpart. Originally, our attention was called to \ks\ in
the course of a cross-identification exercise between X-ray
binaries from the \cite{liu01} catalogue and radio sources from
the Multi-Array Galactic Plane Imaging Survey (MAGPIS) at 6 cm
(\cite{hel06}). While \ks\ is not catalogued as a radio-emitting
X-ray binary, it was intriguing that the MAGPIS radio source
$359.558-0.077$, with a 6 cm peak flux density of 5.5 mJy
beam$^{-1}$, was remarkably consistent with all error boxes listed
in Table \ref{Detections}. Despite the minor coordinate
difference, the MAGPIS object is most likely the same radio source
originally quoted by \cite{ch1994}. From this starting point, we
conducted a follow up investigation whose results are presented in
the following sections.

\begin{table*}
\caption[]{\label{Detections} History of X-ray/low energy $\gamma$-ray detections of KS 1741$-$293 in the literature}
\begin{tabular}{lccccccc}
\hline
\hline
Instrument & Energy range  &  Flux  & Spectral info &  $N_{\rm H}$ & $\alpha_{\rm J2000}$, $\delta_{\rm J2000}$  &  90\% error  & Ref. \\
           &    (keV)      & (erg s$^{-1}$ cm$^{-2}$)  &               &      (cm$^{-2}$)            &      &  radius ($^{\prime\prime}$)  &      \\
\hline
KVANT    &   2-30     & $96 \times 10^{-11}$              & $kT=9$ keV             & $\leq 10^{23}$                                 & $17^h44^m$49\rl 25                   & 60    & 1   \\
         &            &                                 &                        &                                                & $-29^{\circ}21^{\prime}$06\pri 3     &       &     \\
\hline
BeppoSAX &   2-10     & $(14.5_{-0.6}^{+2.0}) \times  10^{-11}$ & $\Gamma = 2.0 \pm 0.2$ & $(20 \pm 2) \times 10^{22}$            & $17^h44^m$49\rl 25                   & 60    & 2     \\
         &            &                                         &                        &                                        & $-29^{\circ}21^{\prime}$06\pri 3     &       &     \\
\hline
ASCA     &   0.7-10   & $9.1 \times 10^{-11}$           & $\Gamma = 2.12_{-0.19}^{+0.20}$ & $(20.3_{-1.6}^{+1.7}) \times 10^{22}$ & $17^h44^m$53\rl 49                   & 40    & 3   \\
         &            &                                 &                        &                                                & $-29^{\circ}21^{\prime}$15\pri 9     &       &     \\
\hline
INTEGRAL &   20-40    & $(6.66\pm0.08) \times 10^{-11}$ &        $-$             &           $-$                                  & $17^h44^m$52\rl 8                         & 64.4  & 4,5 \\
         &   40-100   & $(7.35\pm0.19) \times 10^{-11}$ &                        &                                                & $-29^{\circ}20^{\prime}24^{\prime\prime}$ &       &     \\
\hline
CHANDRA  &   0.5-8    & $2.1 \times 10^{-10}$           &        $\Gamma = 2$      &           $-$                         & $17^h44^m$51\rl 06                        & 1.3  & 6,7 \\
         &            &                                 &
         assumed     &          & $-29^{\circ}21^{\prime}$16\pri 8     &       &     \\
\hline \hline
\end{tabular}
~~\\
(1) \cite{inz1991}; (2) \cite{sid1999}; (3) \cite{sk2002}; (4)
\cite{bel2006}; (5) \cite{brd2006}; (6) \cite{muno06}; (7) this
paper
\end{table*}



\section{\ks\ in the radio}

\begin{table}
\caption[]{\label{vlaobs} VLA archive observations used in this paper \label{obslog}}
\begin{tabular}{cccccc}
\hline
\hline
Date  &  $\lambda$   & VLA    & IF\# and       & \#    &  Project     \\
      &    (cm)      & conf.  &  width$^*$    &  vis. &  id.         \\
\hline
1984 Apr 10 &  6 &  C & 2, 25 & 5245  & A044  \\
2003 Feb 25 &  6 &  D & 2, 50 & 34088 & AL587 \\
2003 Mar 10 &  6 &  D & 2, 50 & 39283 & AL587 \\
2004 Jan 30 & 20 & BC & 2, 50 & 50691 & AL617 \\
2004 Jan 31 & 20 & BC & 2, 50 & 88518 & AL617 \\
2004 May 27 & 20 & CD & 2, 50 & 29511 & AL617 \\
2004 Aug 23 & 20 & D  & 2, 50 & 30897 & AL617 \\
\hline
\hline
\end{tabular}
(*) Frequency width given in MHz.
\end{table}

%
%

   \begin{figure}
   \centering
\resizebox{\hsize}{!}{\includegraphics{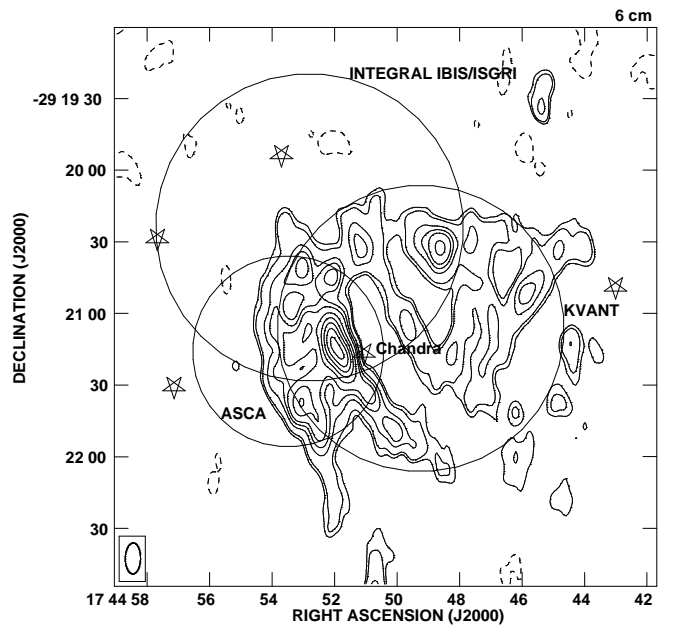}}
      \caption{VLA map of the \ks\ field at the 6 cm wavelength using visibilities
obtained in C and D configurations of the array. Contours are
$-4$, 4, 5, 8, 10, 12, 16, 20, 24, 28, and 32 times 0.4 mJy
beam$^{-1}$, the rms noise. The synthesized beam is shown as an
ellipse at the bottom left corner measuring 12\pri 96 $\times$
6\pri 37 with position angle of $-1^{\circ}$. The 90\% confidence
error circles of the KVANT, ASCA, and INTEGRAL IBIS/ISGRI are also
plotted and labeled accordingly. The positions of Chandra
sources in the field are indicated by the small star symbols whose
uncertainties are smaller than the symbol size. The Chandra label
is reserved for the proposed X-ray counterpart of \ks.}
      \label{vla6cm}
   \end{figure}

   \begin{figure}
   \centering
\resizebox{\hsize}{!}{\includegraphics{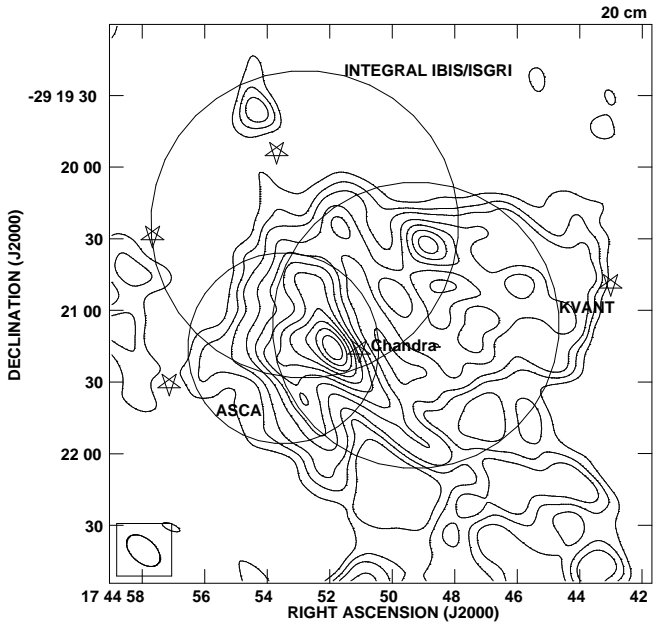}}
      \caption{The same KS~1741$-$293 field and labels as in Fig. \ref{vla6cm} observed
with the VLA at 20 cm. This map has been computed by combining visibilities
obtained in the BC, CD, and D configurations of the array.
Contours are $-4$, 4, 5, 6, 8, 10, 12, 14, 16, 18, 20, and 22 times
1.5 mJy beam$^{-1}$.  The synthesized beam is shown as an ellipse at the bottom left corner
and corresponds to 16\pri 54 $\times$ 9\pri 95 with position angle of 48\grp 6.
}
      \label{vla20cm}
   \end{figure}

%

   \begin{figure}
   \centering
\resizebox{\hsize}{!}{\includegraphics{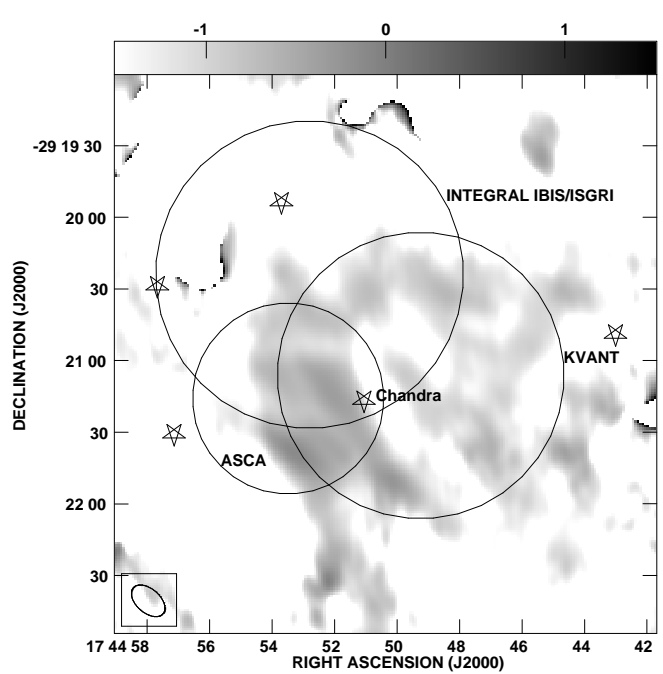}}
      \caption{Spectral index map in gray scale made combining 
the same 6 and 20 cm visibility data of Figs. \ref{vla6cm} and \ref{vla20cm}.
The individual maps were restored with the same 20 cm synthesized beam
(i.e., the one with lower angular resolution) and primary beam corrected prior
to being combined in spectral index mode with the AIPS task COMB.
Labels are the same as in previous figures.
}
      \label{spix}
   \end{figure}


We searched and downloaded several Very Large Array (VLA)
observation projects from the NRAO data archive
covering the \ks\ position at the 6 and 20 cm wavelengths, as listed in Table \ref{vlaobs}. The idea was to produce
radio maps at different wavelengths with similar synthesized beams for reliable spectral
index measurement. The data were edited and calibrated using standard procedures
with the AIPS package of NRAO. In some data sets, the pointing position was slightly different
and the visibilities had to be corrected for the offset. This was achieved using the AIPS task UVFIX
before data sets being concatenated into a single archive with the AIPS task DBCON.
Such a procedure is likely to induce some minor errors when correcting for primary beam response,
but we estimate this should not affect the conclusions based on this work.

The final maps were produced with the AIPS task IMAGR using
uniform weight for enhanced angular resolution. We present them in
Figs. \ref{vla6cm} and \ref{vla20cm} together with the 90\%
confidence error circles of Table \ref{Detections}. The
location of Chandra X-ray sources, that will later be relevant for
the discussion, are also plotted on these maps. It is remarkable
that a peak of radio emission (centered at $\alpha_{\rm
J2000.0}=17^h 44^m$51\rl 96$\pm$0\rl 02 and $\delta_{\rm
J2000.0}=-29^{\circ} 21^{\prime}$14\pri 8$\pm$0\pri 4) is clearly
visible within the overlapping region of the KVANT, ASCA, and
INTEGRAL IBIS/ISGRI circles. This maximum, together with the
nearby extended radio emission, are very likely to conform the
arc-minute extended radio source quoted by \cite{ch1994} based on
single-dish observations. A spectral index map shown in Fig.
\ref{spix} was computed by combining matching beam images made
from the same visibility data of Figs. \ref{vla6cm} and
\ref{vla20cm} after correction for primary beam response. As a
result, negative spectral index values significantly dominate the
field, which is indicative of the fact that that the whole radio source is of
non-thermal (i.e., synchrotron) origin.

\section{\ks\ in X-rays}

The recent Chandra Galactic Central 150 pc Source Catalogue (\cite{muno06})
shows several X-ray sources in the \ks\ field. Only one of them,
namely CXOGC J174451.0$-$292116, is consistent with the overlapping region
of all error circles plotted in previous figures. This fact certainly appears
remarkable and leads us to propose this Chandra source as the most likely X-ray counterpart candidate to \ks.
The Chandra detection appears to have a negligible count rate in the softest energy band of the satellite.
This fact is well consistent with a highly absorbed source.


\section{\ks\ in the infrared and optical}

\begin{figure*}
\mbox{}
\resizebox{\hsize}{!}{\includegraphics{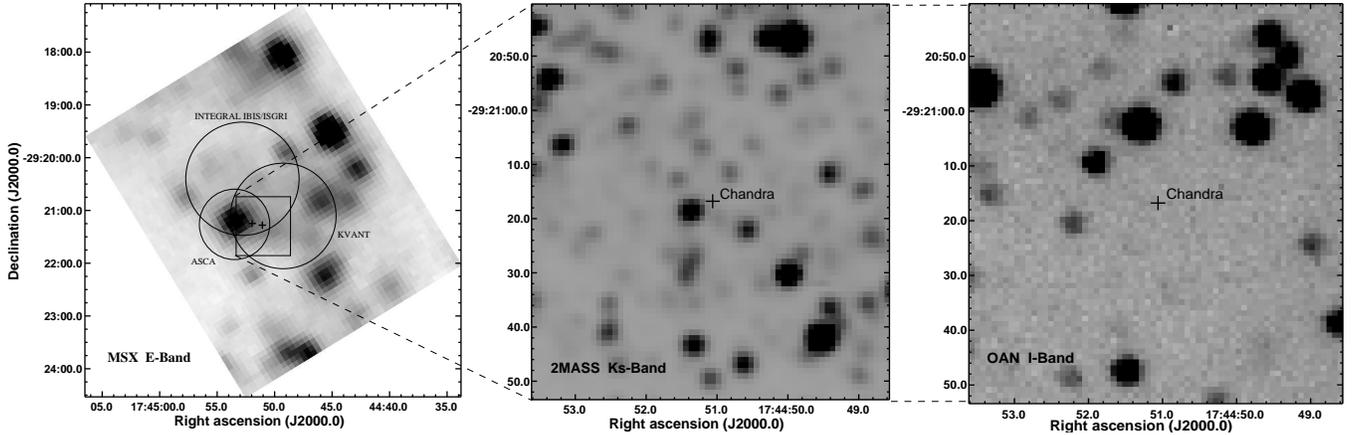}}
\caption{
{\bf Left.} The field of \ks\ as observed with MSX in the infrared E-band (21.3 $\mu$m).
The same KVANT, ASCA, and INTEGRAL IBIS/ISGRI error circles are plotted as in previous figures.
The small left and right central crosses mark the location of the radio peak
and the Chandra X-ray source discussed in the text, respectively.
{\bf Middle.} Zoomed 2MASS image in the near infrared $Ks$-band centered around the proposed Chandra counterpart of
KS~1741$-$293. The location and uncertainty
of the Chandra X-ray source is indicated by the small cross.
{\bf Right.} The same as in the middle panel observed in the optical $I$-band.
This frame was obtained on August 28th, 2006, with the OAN 1.52~m telescope
in Calar Alto (Spain).
}
\label{panels}
\end{figure*}

In the left panel of Fig. \ref{panels} we show the 21.3 $\mu$m image of the \ks\ region
as observed by the NASA Midcourse Space Experiment (MSX). Here, there is also
an obvious source, of point-like appearance, within the common area of error circles. It is
named as G359.5609$-$00.0810 in the MSX catalogue and is located at
$\alpha_{\rm J2000.0}=17^h 44^m$53\rl 28$\pm$0\rl 03 and
$\delta_{\rm J2000.0}=-29^{\circ} 21^{\prime}$10\pri 8$\pm$0\pri 2.
The rising flux density of this mid-infrared source across the MSX bands
is suggestive of being a highly absorbed object. Its position is within the large uncertainty
of the far-infrared source FIR5 proposed by \cite{ch1994}. However,
looking at the original FIR5 data (\cite{fir5}), one cannot rule out
confusion problems when trying to associate it with a single MSX source.
A key point here is the fact that the MSX source G359.5609$-$00.0810
is not consistent in position with the peak of radio emission in the VLA maps,
suggesting that they could be unrelated objects.

The Chandra position has a 90\% confidence error circle of 1\pri 3 thus enabling
accurate searches for counterparts at other wavelengths that could confirm the identification.
A preliminary inspection of the  2 Micron All Sky Survey (2MASS, \cite{st2006}) images
and a deep CCD optical integration reveal no near infrared or optical counterpart
for CXOGC J174451.0$-$292116 (see middle and right panels of Fig. \ref{panels}).
This is not a surprising result given the large absorption
expected in this field. The currently available magnitude limits are about
$I \geq 20.5$, $J \geq 18$, $H \geq 15$, and $Ks \geq 14$.

\section{Discusion and conclusions}

Our main argument to claim the identification of \ks\ with the Chandra source CXOGC J174451.0$-$292116
is a positional coincidence with all error circles available.
The unabsorbed flux from CXOGC J174451.0$-$292116, in the 0.5-8 keV band, is relatively faint for
accurate spectral analysis and amounts to $4.1 \times 10^{-6}$ photon cm$^{-2}$ s$^{-1}$.
Assuming a conceivable photon index $\Gamma=1.5$ and hydrogen
column density $N_H = 1.0 \times 10^{23}$ cm$^{-2}$, the corresponding
unabsorbed energy flux in the 2-10 keV band is $9.6 \times 10^{-10}$ erg s$^{-1}$ cm$^{-2}$.
At a Galactic center distance of 8.5 kpc, this corresponds to a luminosity
of $8.3 \times 10^{36}$ erg s$^{-1}$, which appears as a reasonable value for the quiescent
emission of a X-ray binary.
For the hydrogen column density adopted above, the deepest $Ks$-band upper limit would translate
into absolute magnitudes of $Ks_{\rm abs} \geq -6$ at the Galactic center distance.
This value safely excludes at least all supergiant and late giant stars and would be fully in agreement with
the suspected LMXB nature of \ks.

If our identification is correct, any association with the
\cite{ch1994} radio and far-infrared is definitely ruled out. This
is because CXOGC J174451.0$-$292116 is not positionally consistent
with either the radio peak or the MSX source discussed above. On
the other hand, the fact the Chandra source itself is not
detected at 6 cm above 1.2 mJy (3$\sigma$) is compatible with the
usually low radio luminosities of neutron star X-ray binaries
(\cite{muno05}).

The morphology of the extended radio emission around the Chandra source is very reminiscent of a
supernova remnant (SNR). This is also supported by the non-thermal spectral index estimated from 20 and 6 cm maps.
Therefore an interesting scenario is that this extended radio source, first pointed out by
\cite{ch1994}, could be the SNR where \ks\ was originated.

The association between LMXBs with SNRs could be the result of supernova
explosions due to the accretion induced collapses, a really rare event.
Searching for LMXBs projected on SNRs in the Galaxy, \cite{tuncer00} found that of six positional
coincidences only one of them is likely to be real. This is the case of the LMXB
2259+587 and the SNR G109.1-1.0 (CTB 109) (\cite{gf80}; \cite{hs94}), a system that
lies at 3 kpc. In our case, adopting a distance for the possible
LMXB/SNR system of $\sim8.5$ kpc, the
SNR radius should be of 3.8 pc. If the expansion is adiabatic, it can be
modeled by the standard Sedov solutions (\cite{sedov59}). Considering the
SNR expansion in a homogeneus medium with density $\sim 1$ cm$^{-3}$ around the
galactic center and an SN energy of the explosion $E \sim 10^{51}$ erg
(\cite{spitzer04}), the age of the SNR would be $t \sim 500$ yr.
This age is very low for a LMXB, so that the association of \ks\
with the SNR does not seem very likely unless the source is very young.

In summary, we have revisited the previous radio and mid-infrared associations for \ks\ showing that they are not
sustained by the modern observational data. Instead, we propose a Chandra X-ray source as a likely counterpart.
Our candidate is surrounded by extended non-thermal radio emission,
with a very likely SNR origin, although a physical association is not yet clear.
Further progress on unveiling the true nature of \ks\ is needed given that it is one of the
conspicuous objects in the Galactic center that also appears superposed to the TeV emission from the Galactic Ridge
(\cite{aha2006}). This could be achieved by deeper infrared observations.

\begin{acknowledgements}
{\small The authors acknowledge support by
grant AYA2004-07171-C02-02 from the Spanish government, FEDER
funds, and FQM322 of Junta de Andaluc\'{\i}a.
This research made use of the SIMBAD
database, operated at the CDS, Strasbourg, France.
J.A.C. is a researcher in the programme {\em Ram\'on y Cajal} funded
jointly by the Spanish Ministerio de
Ciencia y Tecnolog\'{\i}a and Universidad de Ja\'en.
D.P.R also acknowledges Junta de Andaluc\'{\i}a (Spain).
The NRAO is a facility of the NSF
operated under cooperative agreement by
Associated Universities, Inc. in the U.S.A.
This research has made use of the NASA/ IPAC Infrared Science Archive,
which is operated by the JPL
CIT,  under contract with the
NASA.  This publication makes use of data products from the
2MASS,  which is a joint project
of the University of Massachusetts and the Infrared
Processing and Analysis Center/CIT,
funded by NASA
and the NSF, in the U.S.A. We also thank the Observatorio Astron\'omico Nacional (OAN) staff of the
1.52~m telescope in Calar Alto (Spain).
}
\end{acknowledgements}


\end{document}